\documentclass[aps, prl, twocolumn, superscriptaddress, showpacs]{revtex4}
\usepackage{mathrsfs, amsfonts, amsmath, amssymb, graphicx, bm, epsfig}
\usepackage[colorlinks, linkcolor=blue, anchorcolor=blue, citecolor=red]{hyperref}

\begin{document}
\title{Dynamics of Antiferromagnets Driven by Spin Current}
\author{Ran Cheng}
\email{rancheng@utexas.edu}
\affiliation{Department of Physics, University of Texas at Austin, Austin, Texas 78712, USA}
\author{Qian Niu}
\affiliation{Department of Physics, University of Texas at Austin, Austin, Texas 78712, USA}
\affiliation{International Center for Quantum Materials, Peking University, Beijing 100871, China}

\begin{abstract}
When a spin-polarized current flows through a ferromagnetic (FM) metal, angular momentum is transferred to the background magnetization via spin-transfer torques. In antiferromagnetic (AFM) materials, however, the corresponding problem is unsolved. We derive microscopically the dynamics of an AFM system driven by spin current generated by an attached FM polarizer, and find that the spin current exerts a driving force on the local staggered order parameter. The mechanism does not rely on the conservation of spin angular momentum, nor does it depend on the induced FM moments on top the AFM background. Two examples are studied: (i) A domain wall is accelerated to a terminal velocity by purely adiabatic effect where the Walker's break-down is avoided; and (ii) Spin injection modifies the AFM resonance frequency, and spin current injection triggers spin wave instability of local moments above a threshold.
\end{abstract}

\maketitle

\textit{Introduction.}---Mutual dependence of current and magnetization is the central problem of spintronics~\cite{ref:spintronics}, which can be described in a complementary way. In a ferromagnetic (FM) material where local magnetization varies slowly over space and time, conduction electron spins will follow the orientation of the background magnetization, known as the adiabatic limit~\cite{ref:Volovik}. In turn, spin angular momentum is transferred to the background magnetization via spin transfer torques~\cite{ref:BegerSlonczewski,ref:Bazaliy,ref:STT,ref:STTReview1,ref:Tserkovnyak,ref:STTReview2} as a result of the conservation of spin angular momentum. Spin transfer torques provide key mechanisms to many intriguing phenomena in FM materials such as current-driven domain wall dynamics~\cite{ref:DWDynamics1,ref:DWDynamics2}, spin wave excitations~\cite{ref:SW1,ref:SW2,ref:SW3}, \emph{etc.}. However, in antiferromagnetic (AFM) materials, corresponding issues are unsolved puzzles hindered by two fundamental difficulties: (i) staggered AFM order does not respect spin conservation with conduction electrons; (ii) neighboring magnetic moments are anti-parallel so that the adiabatic electron dynamics derived in FM materials no longer applies.

On the other hand, many recent experiments~\cite{ref:Tsoi,ref:Spincurrent,ref:AFMSV} and numerical simulations~\cite{ref:Numerical1,ref:Numerical2,ref:Numerical3} indicate that AFM materials exhibit current-induced effects with similar orders of magnitude, if not stronger than, as those in ferromagnets. Those pioneering investigations ushered the field of AFM spintronics~\cite{ref:AFMSpintronics} and propelled AFM materials as promising candidates for real applications. From a theoretical point of view, AFM dynamics driven by charge current has been studied both phenomenologically~\cite{ref:Brataas1,ref:Brataas2} and microscopically~\cite{ref:Duine,ref:MacD}. In the former, both adiabatic torque by \textit{ac} current and non-adiabatic torque by \textit{dc} current are predicted, but an adiabatic effect in the \textit{dc} limit is absent; in the latter, adiabatic torque is generated by \textit{dc} current, but the result includes only second-order derivatives in space and time. Case becomes rather unclear when turning to spin current, which can be realized by attaching a FM polarizer to the system. This problem has only been explored phenomenologically~\cite{ref:Kunitsyn} and no microscopic study is yet available. Even in the phenomenological model, it is the induced FM moments on top of the AFM background that respond to the spin current, which is a higher order effect that drives the AFM staggered order \textit{indirectly}. Will a spin current respond to and drive the staggered order \textit{directly} without the help of induced FM moments?

We have answered part of this question in a previous publication~\cite{ref:rancheng}, where the adiabatic dynamics of conduction electrons is developed in an AFM material with given background profile. In this Letter, we solve the converse --- how a spin current exerts back-action on the AFM background. In analogy to ferromagnets, electron dynamics becomes adiabatic when the AFM staggered order parameter $\bm{n}=(\bm{m}_A-\bm{m}_B)/2$ ($\bm{m}_A$ and $\bm{m}_B$ are neighboring magnetic moments) is slowly varying~\cite{ref:rancheng}. However, instead of following the background strictly, electrons are subject to internal dynamics between degenerate bands, which results in mistracking with the background even in the adiabatic limit. The underlying physics is that the anti-parallel moments introduce an internal degree of freedom on conduction electrons which absorbs dynamics \textit{within} a unit cell, while dynamics \textit{among} unit cells is governed by the slowly-varying $\bm{n}(\bm{r},t)$~\cite{ref:note}. We will follow the same idea but our target here is the equation of motion for $\bm{n}(\bm{r},t)$.

\textit{Formalism.}---We adopt the Lagrangian approach where the system Lagrangian L is
\begin{align}
 L=\int\mathrm{d}^{d}r\mathcal{L}=\int\mathrm{d}^{d}r(\mathcal{L}_n+\mathcal{L}_{int}), \label{eq:L}
\end{align}
with $d$ being the dimensionality. $\mathcal{L}_n$ and $\mathcal{L}_{int}$ represent Lagrangian densities of the AFM background and its interaction with conduction electrons, respectively. When $\bm{n}(\bm{r},t)$ is slowly varying, the AFM background is effectively described by the non-linear $\sigma$ model~\cite{ref:Haldane}
\begin{align}
 \mathcal{L}_n=\frac1{2g}[\frac1c(\partial_t\bm{n})^2-c|\nabla\bm{n}|^2-\frac{\omega_0^2}{c}\bm{n}_{\perp}^2], \label{eq:Ln}
\end{align}
where $c=2aSJ/\hbar$ denotes the spin wave velocity, $g=2\sqrt{d}a^{d-1}/\hbar S$ is the coupling coefficient with $a$ being the lattice constant and $S$ being the spin of core magnetic moments. The last term in Eq.~\eqref{eq:Ln} describes uniaxial anisotropy, where $\bm{n}_{\perp}$ includes components of $\bm{n}$ perpendicular to the easy axis.

The interaction term $\mathcal{L}_{int}$ is constructed by summing over contributions from individual electrons: $\mathcal{L}_{int}=\sum_{\lambda}\int\mathrm{d}^d\bm{k}L_{\lambda}(\bm{k})f_{\lambda}(\bm{k})$, where $L_{\lambda}(\bm{k})$ is the Lagrangian of an electron with momentum $\bm{k}$ in band $\lambda$, and $f_{\lambda}(\bm{k})$ is the distribution function. As was shown in Ref.~\cite{ref:rancheng}, a slowly-varying $\bm{n}(\bm{r},t)$ in space-time yields an effective description of electron dynamics in terms of the Berry phase theory~\cite{ref:DiXiao}, where the influence of AFM background on a single electron is recast into the coupling of a Berry gauge field. This gauge field has an important counterpart in FM materials, which is responsible for the spin-motive force~\cite{ref:SMF,ref:Shengyuan} and the topological Hall effect~\cite{ref:THE}. Using the same gauge field in FM textures, the adiabatic spin-transfer torque (or reactive torque) can be derived as a complementary effect~\cite{ref:Volovik,ref:Bazaliy}. What we shall do here is a parallel job in AFM textures: seeking the complementary effect of Ref.~\cite{ref:rancheng} using the Berry gauge field derived there. The variational derivative of the Berry phase term of $L_{\lambda}(\bm{k})$ with respect to $\bm{n}(\bm{r},t)$ gives~\cite{ref:supplementary}
\begin{align}
 \frac{\delta L_{\pm}(\bm{k})}{\delta\bm{n}}=\pm\frac{\hbar}2(1-\xi^2)\bm{n}\times[\partial_t\bm{n}+\bm{v}_e(\bm{k})\cdot\nabla\bm{n}], \label{eq:deltaLe}
\end{align}
where $\pm$ distinguishes two sub-bands that are degenerate in energy, and $\bm{v}_e(\bm{k})=\frac1{\hbar}\frac{\partial\varepsilon}{\partial\bm{k}}$ is the band velocity of the electron. The parameter $\xi$ is a constant of motion derived before~\cite{ref:rancheng} but its specific form is not important here. From Eq.~\eqref{eq:deltaLe}, we know
\begin{align}
 \frac{\delta\mathcal{L}_{int}}{\delta\bm{n}}&=\sum_{\lambda}\int\mathrm{d}^d\bm{k}\frac{\delta}{\delta\bm{n}} L_{\lambda}(\bm{k})f_{\lambda}(\bm{k}) \notag\\
 &=\frac{\hbar}{2}(1-\xi_F^2)\bm{n}\times[\rho_s\frac{\partial\bm{n}}{\partial t}+(\bm{j}_s\cdot\nabla)\bm{n}], \label{eq:delaLint}
\end{align}
where $\xi_F$ is the value of $\xi$ at Fermi energy; $\rho_s$ and $\bm{j}_s$ are the spin density and the spin current density \textit{with respect to the local order parameter $\bm{n}(\bm{r},t)$}~\cite{ref:supplementary}. There are many possible ways to inject spin current into antiferromagnets: for example, attaching a FM polarizer, spin pumping~\cite{ref:spinpumping}, and spin photovoltaic effect~\cite{ref:photovoltaic}. But in the following, we will focus on the first case; the FM polarizer is assumed to be a half metal for estimations, but we will comment on the case of an imperfect polarizer. In this case, $\rho_s$ and $\bm{j}_s$ are both proportional to the current density $\bm{j}_c$, thus the ratio $\bm{v}_s\equiv\bm{j}_s/\rho_s$ is independent of $\bm{j}_c$. In typical collinear AFM metals~\cite{ref:material1,ref:material2}, $v_s$ is estimated to be $10^5$~cm/s at room temperature~\cite{ref:supplementary}. As will be explained in a separate publication, spin-flip processes are highly suppressed by the smallness of $\xi_F$, for which AFM metals are better spin-preservers compared to normal metals. Thus we will approximate $\rho_s$ and the magnitude of $\bm{j}_s$ as spatially uniform.

To derive the equation of motion, we further need to account for Gilbert damping by the Rayleigh's dissipation function $R=\int\mathrm{d}^{d}r\mathcal{R}=\alpha\int\mathrm{d}^{d}r\dot{\bm{n}}^2$ where $\alpha$ is phenomenological. The full Lagrangian equation becomes:
$\frac{\partial\mathcal{L}}{\partial\bm{n}}-\frac{\mathrm{d}}{\mathrm{d}t}\frac{\partial\mathcal{L}}{\partial\dot{\bm{n}}}- \nabla\!\cdot\!\frac{\partial\mathcal{L}}{\partial(\nabla\bm{n})}=\frac{\partial\mathcal{R}}{\partial\dot{\bm{n}}}$. It allows us to obtain the central result in view of the constraint $\bm{n}^2=1$~\cite{ref:supplementary}:
\begin{align}
 \bm{n}\times[\partial_t^2\bm{n}-c^2\nabla^2\bm{n}&+\omega_0^2\bm{n}_{\perp}]+\tilde{\alpha}\bm{n}\times\partial_t\bm{n}\notag\\
 &+\mathcal{G}(\rho_s\partial_t+\bm{j}_s\cdot\nabla)\bm{n}=0, \label{eq:eom}
\end{align}
where $\mathcal{G}=\frac{c\sqrt{d}}{Sa}(1-\xi_F^2)$ denotes the coupling strength, and $\tilde{\alpha}=2\alpha c\sqrt{d}a^{d-1}/\hbar S$ is the effective damping coefficient. Two special features of Eq.~\eqref{eq:eom} deserve emphasis: (i) Under time reversal operation, $\bm{n}$ flips, but the polarizer is also reversed, thus $\rho_s$ is kept the same while $\bm{j}_s$ changes sign (illustrated by Fig.~1 of Ref.~\cite{ref:supplementary}). Therefore, all terms in Eq.~\eqref{eq:eom} respect the same time reversal symmetry except the damping term. This is consistent with our adiabatic assumption, since adiabatic terms are usually non-dissipative. (ii) Though similar to the adiabatic torque in ferromagnets, the term $\bm{j}_s\cdot\nabla\bm{n}$ does not behave as a torque, it is a driving force since the AFM dynamics is second order in time derivative.

\begin{figure}[t]
   \centering
   \includegraphics[width=0.9\linewidth]{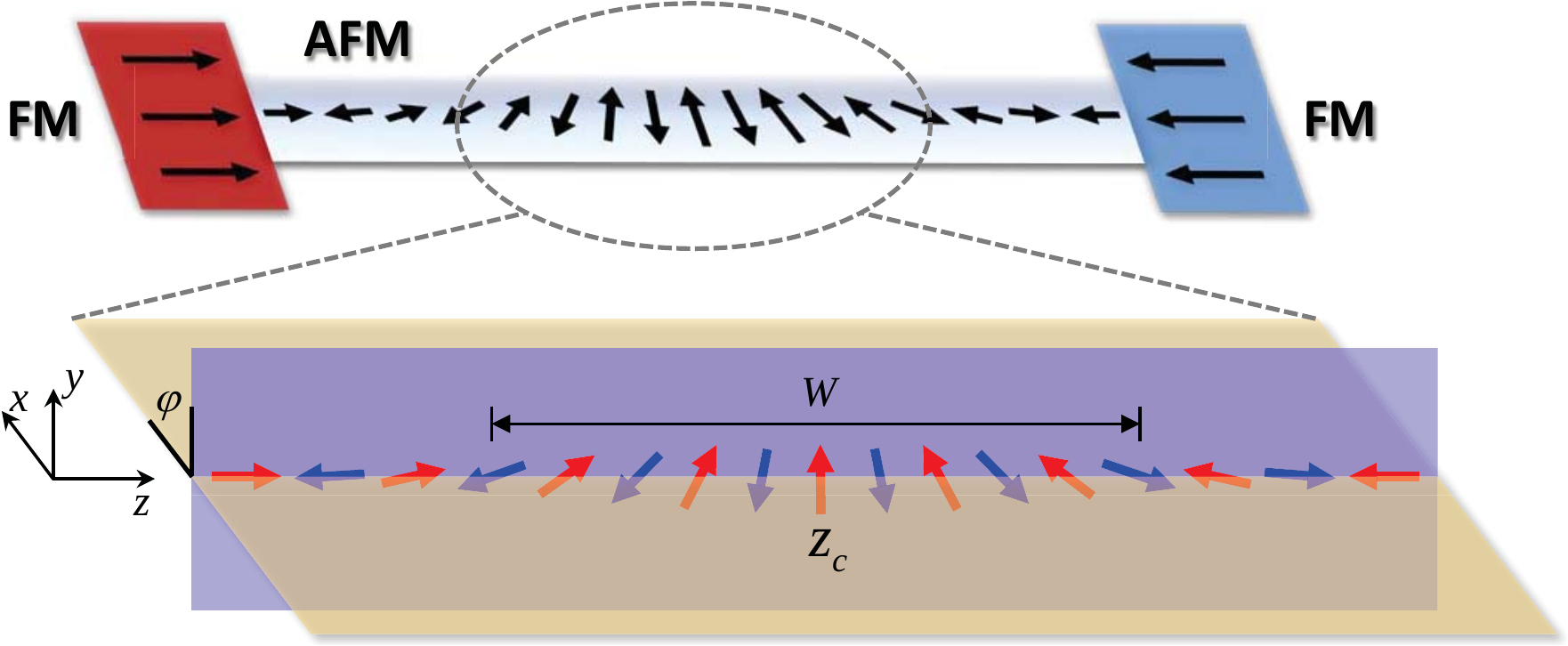}
   \caption{(Color online) Schematic view of a setup of AFM DW between two pinning ferromagnets at its ends. DW dynamics is described by two collective coordinates, the center position $z_c$ and the canting angle $\varphi$. The DW width $W$ is approximately invariant during the motion.}{\label{Fig:Device}}
\end{figure}

\textit{Comparisons}.---We compare Eq.~\eqref{eq:eom} with results from existing literature. Ref.~\cite{ref:Brataas1,ref:Brataas2} studied charge current effects, but within the adiabatic limit and linear order in $\nabla\bm{n}$, only \textit{ac} current has non-trivial consequence, whereas \textit{dc} current produces null result. In contrast, $\rho_s\partial_t\bm{n}$ and $\bm{j}_s\cdot\nabla\bm{n}$ are \textit{linear order} adiabatic terms in the \textit{dc} limit. Ref.~\cite{ref:Duine} studied \textit{dc} charge current, but the result contains only second order terms $\partial_t^2\bm{n},\nabla^2\bm{n}$, and $\partial_t\nabla\bm{n}$. Ref.~\cite{ref:Kunitsyn} considered spin current, but it only couples to the induced FM moments on top of the AFM background, which drives the staggered order indirectly, and the result is again second order.

\begin{figure}[t]
   \centering
   \includegraphics[width=0.83\linewidth]{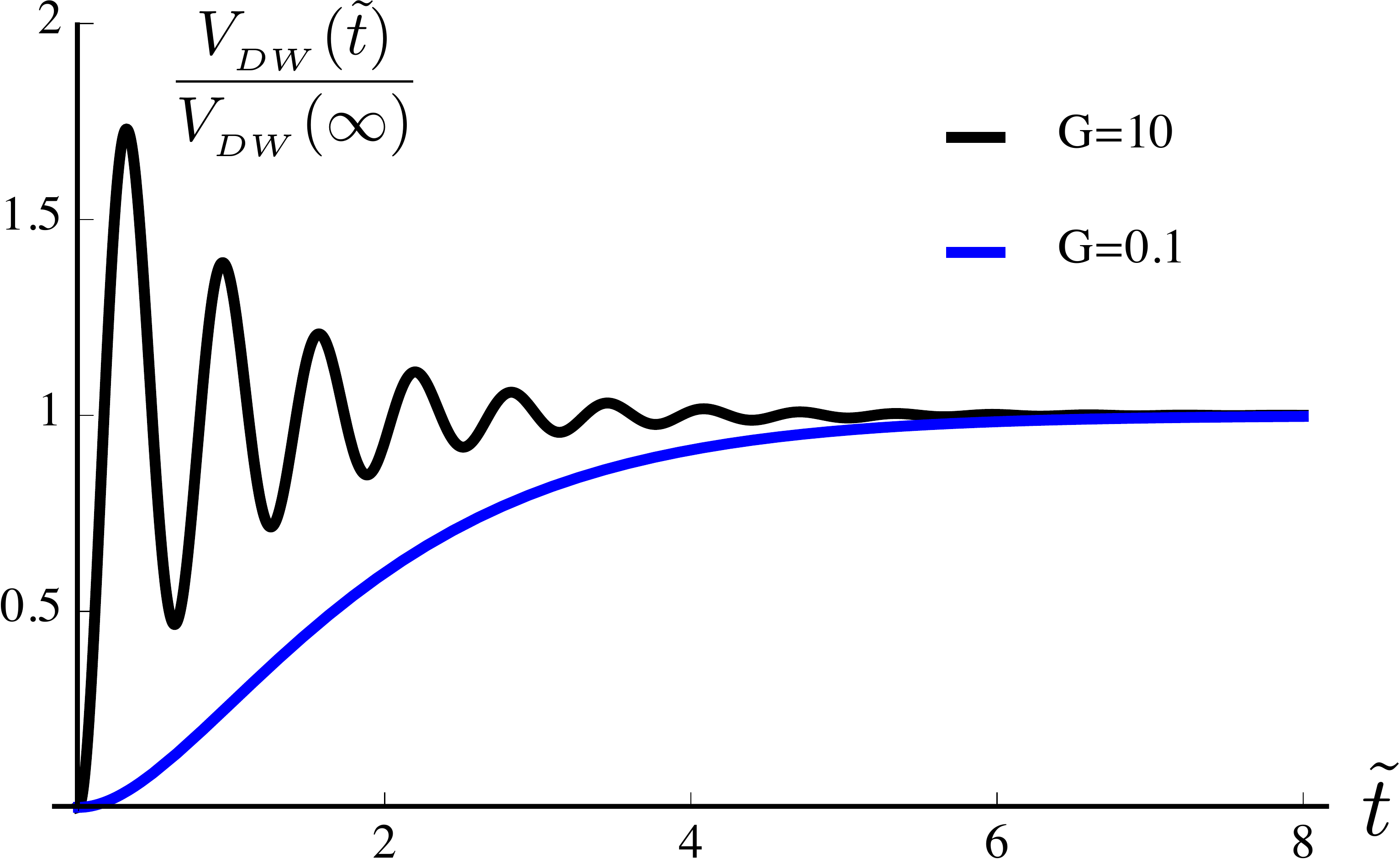}
   \caption{(Color online) Scaled DW velocity plotted as a function of time, for $G=0.1$ and $G=10$, respectively. $V_{_{DW}}$ exhibits damped oscillation with $V_{_{DW}}(\infty)=G^2/(1+G^2)$.}{\label{Fig:DWvelocity}}
\end{figure}

\emph{Domain Wall Dynamics.}---Due to the absence of dipolar interaction, formation of an AFM domain wall (DW) requires two pinning ferromagnets (along the easy axis) at the ends. The pinning originates from exchange bias effect on the interface between FM and AFM materials~\cite{ref:ExchBias}. Consider the DW of 180 degree depicted in Fig.~\ref{Fig:Device}; such a texture can be achieved by first growing two pinning FM layers on a homogeneous AFM metal, then rotating one of them to the opposite direction. Though not in exact agreement with theoretical prediction~\cite{ref:DW}, it has been realized experimentally in many different contexts~\cite{ref:Chien,ref:Bode}. As a compromise between exchange interaction and anisotropy, the DW assumes a soliton profile~\cite{ref:DW}. When the DW is moving, we describe it by the Walker's ansatz~\cite{ref:Walker}:
\begin{align}
 \varphi(z,t)=\varphi(t);\quad \tan\frac{\theta(z,t)}2=\exp\left[\frac{z-z_c(t)}{W(t)}\right], \label{eq:Walker}
\end{align}
where $\varphi$ and $\theta$ are spherical angles specifying the local orientation of $\bm{n}(\bm{r},t)$. The first equation states that $\bm{n}$-vectors at different positions are kept coplanar and have a common canting angle. The second equation implies that the DW remains a soliton shape except that its width $W(t)$ varies with time and that the DW moves as a whole with an instantaneous center position $z_c(t)$. Eq.~\eqref{eq:Walker} enables us to compute the total Lagrangian as a function of three parameters $z_c$, $\varphi$, and $W$, known as the collective coordinates. When DW velocity is much smaller than $c$ and its rotation rate is much lower than $\omega_0$, it can be shown that $W(t)$ is essentially a constant of motion. Hence we are left with only two dynamical variables $z_c$ and $\varphi$. Not bothering with an overall factor, the system Lagrangian is effectively written as
\begin{align}
 L=\frac{\dot{z}_c^2}{W}+W\dot\varphi^2+2\mathcal{G}(\rho_s z_c\dot{\varphi}+j_s\varphi). \label{eq:DWLagrangian}
\end{align}
The Rayleigh's dissipation function can be calculated in a similar way, $R=\tilde{\alpha}(\dot{z}^2_c/W+W\dot{\varphi}^2)$. After some straightforward algebra, we obtain the equations of motion
\begin{subequations}
\label{eq:DWeom}
\begin{align}
 &\ddot{z}_c+\tilde{\alpha}\dot{z}_c=\rho_s\mathcal{G}W\dot{\varphi}, \\
 &\ddot{\varphi}+\tilde{\alpha}\dot{\varphi}=\frac{\rho_s\mathcal{G}}{W}(v_s-\dot{z}_c),
\end{align}
\end{subequations}
which can be solved analytically. To measure time by $\tilde{\alpha}^{-1}$ and the DW center velocity by $v_s$, we define $V_{_{\mathrm{DW}}}\equiv\dot{z}_c/v_s$ and $\tilde{t}\equiv\tilde{\alpha}t$. By eliminating $\varphi$ in Eq.~\eqref{eq:DWeom}, we obtain
\begin{align}
 \ddot{V}_{_{\mathrm{DW}}}+2\dot{V}_{_{\mathrm{DW}}}+(G^2+1)V_{_{\mathrm{DW}}}=G^2, \label{eq:HO}
\end{align}
where $G=\rho_s\mathcal{G}/\tilde{\alpha}$. Eq.~\eqref{eq:HO} describes an underdamped harmonic oscillator driven by a constant force. For the initial condition $V_{_{\mathrm{DW}}}(0)=0$, the solution is
\begin{align}
 V_{_{\mathrm{DW}}}=\frac{G^2-Ge^{-\tilde{t}}[G\cos G\tilde{t}+\sin G\tilde{t}]}{1+G^2}, \label{eq:VDW}
\end{align}
which is plotted in Fig.~\ref{Fig:DWvelocity} for two different $G$'s. As $\tilde{t}\rightarrow\infty$, $V_{_{\mathrm{DW}}}$ terminates at $V_{_{\mathrm{DW}}}(\infty)=G^2/(1+G^2)$. As mentioned before, $\rho_s$ is proportional to the current density $j_c$, and so is $G$. Therefore, $V_{_{\mathrm{DW}}}(\infty)$ is quadratic in $j_c$ for small current and approaches $v_s$ as a limit at extremely large current. However, the DW velocity may not saturate at $v_s$ when effects due to pure charge current are considered~\cite{ref:Brataas1,ref:Brataas2}.

Regarding pure spin current effect alone, we estimate for typical collinear AFM metals, such as IrMn and PdMn~\cite{ref:material1,ref:material2}. The core spin is $2\sim4$~$\mathrm{\mu}_B$; $c$ is of order $10^5$ cm/s; $a$ is $3.6\sim 3.8$~$\mathrm{\AA}$; the damping rate is similar to FM metals thus $\tilde{\alpha}\sim 10^9\ s^{-1}$. For a current density of $10^5$~A/cm$^2$, $G$ is somewhere between $0.1$ and $1$, thus the DW is driven up to $10^4$~cm/s. As a comparison, the same DW velocity in ferromagnets requires $10^8$~A/cm$^2$, which means that an AFM DW is easier to drive. However, if the polarizer is not half metallic, for example, with a polarization of $0.7$, the required current density will be raised up to roughly $10^7$~A/cm$^2$.

To close the argument, three remarks are in order. (i) Eliminating $z_c$ in Eq.~\eqref{eq:DWeom} gives an equation of $\dot{\varphi}$ (not $\varphi$ itself) very similar to Eq.~\eqref{eq:HO}, which indicates that no matter how slow the DW center moves, it is always accompanied by the precession of $\varphi$. This is in sharp contrast to the DW dynamics in ferromagnets, where precession only occurs after the Walker's break-down~\cite{ref:DWDynamics1,ref:DWDynamics2}. What removes the Walker's break-down here is the absence of demagnetization due to vanishing net magnetization. (ii) Our theory is based on adiabatic electron dynamics, thus $\mathcal{G}(\rho_s\partial_t+\bm{j}_s\cdot\nabla)\bm{n}$ only includes the adiabatic effect of spin current. While \textit{only} non-adiabatic torque determines the terminal velocity of a FM DW~\cite{ref:STT}, the AFM DW here is driven to a steady motion by \textit{purely} adiabatic forcing, the transfer efficiency of which is usually much higher than that of non-adiabatic effects. This is responsible for why an AFM DW is more movable. (iii) When a DW is passing by, local moments will be dragged away from the easy axis, which result in a change of anisotropic magnetoresistance along the transverse direction. This provides a possible way to monitor the DW motion.

\begin{figure}[b]
   \centering
   \includegraphics[width=0.85\linewidth]{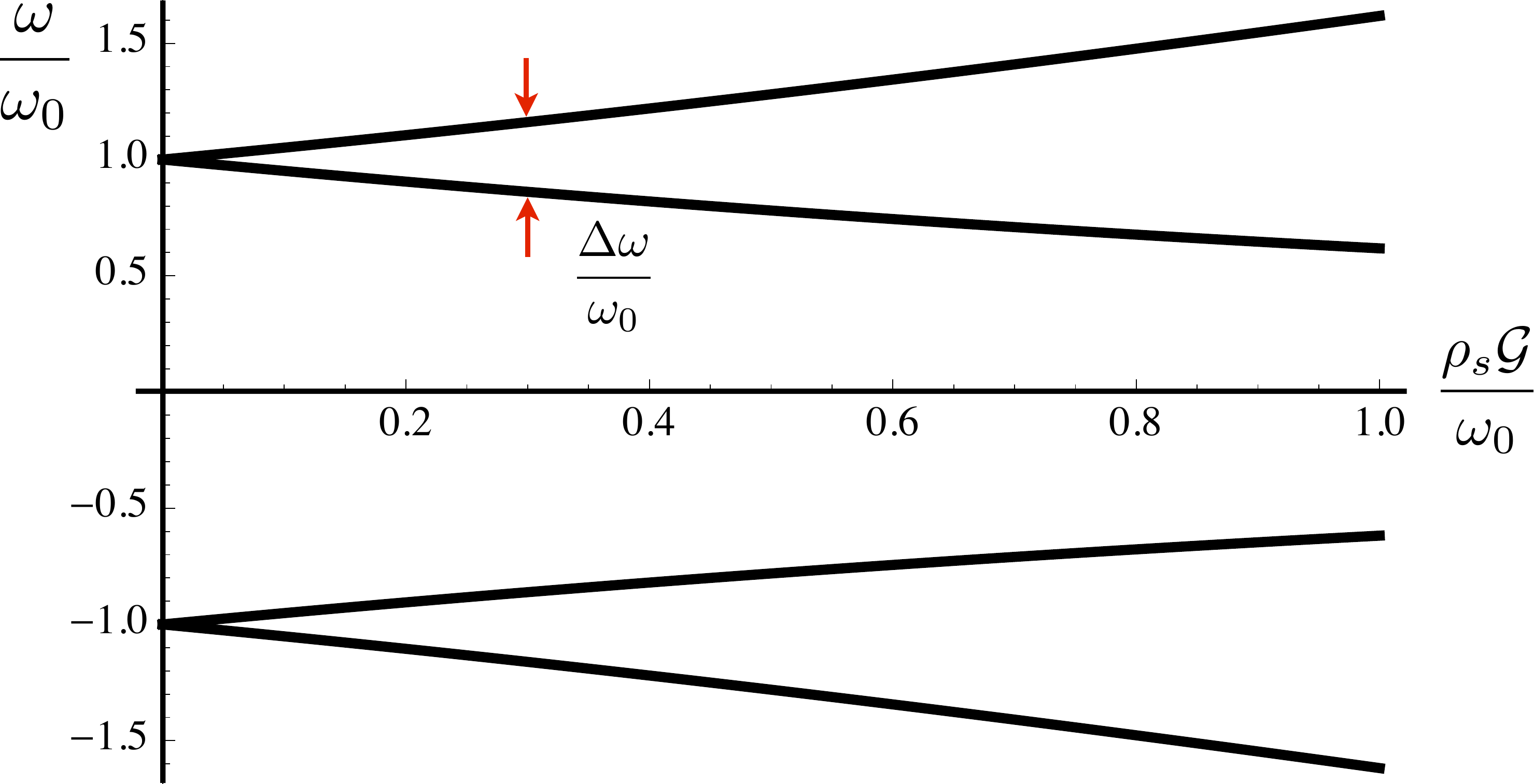}
   \caption{Spin wave spectrum (at zero $k$) as a function of spin injection. As to whether the A or B sublattice is pinned along the FM polarizer, there is a sizable difference in the AFM resonance frequency represented by $\Delta\omega$.
   }{\label{Fig:Splitting}}
\end{figure}

\emph{Spin Wave Excitations.}---Injection of spin current significantly modifies spin wave excitations in antiferromagnets. We take the Ansatz $\bm{n}=\hat{\bm{e}}+\bm{n}_{\perp}e^{i(\bm{k}\cdot\bm{r}-\omega t)}$, where $\bm{n}_{\perp}$ is a small deviation ($|\bm{n}_{\perp}|\ll1$) perpendicular to the easy axis $\hat{\bm{e}}$. It is worth mentioning that the relative motion between $\bm{m}_A$ and $\bm{m}_B$ within a unit cell [the dynamics of $\bm{m}=(\bm{m}_A+\bm{m}_B)/2$ with the constraint $\bm{m}\cdot\bm{n}=0$] seems to have been ignored, but in fact it has been \textit{resolved} into the dynamics of $\bm{n}$ described by Eq.~\eqref{eq:Ln}~\cite{ref:Haldane}. By substituting the above ansatz into Eq.~\eqref{eq:eom}, we obtain
\begin{align}
 (\omega^2-\omega_0^2-c^2k^2)+i\tilde{\alpha}\omega\pm\rho_s\mathcal{G}(\omega-\bm{v}_s\cdot\bm{k})=0, \label{eq:SWSpectrum}
\end{align}
where $+$ ($-$) refers to the case where the direction of the $A$ ($B$) sublattice is pinned along the FM polarizer. First consider the macrospin model that the system precesses as a whole ($k=0$); hence Eq.~\eqref{eq:SWSpectrum} is solved as
\begin{align}
 \mathrm{Re}[\omega]=\frac12[\pm\rho_s\mathcal{G}\pm\sqrt{(\rho_s\mathcal{G})^2+4\omega_0^2}], \label{eq:fourmodels}
\end{align}
where the two $\pm$ are independent. The first $+(-)$ sign represents that the polarizer pins the A (B) sublattice. Eq.~\eqref{eq:fourmodels} is plotted in Fig.~\ref{Fig:Splitting}; we see that the frequency difference $\Delta\omega$ for opposite polarizer orientations is proportional to the spin density $\rho_s$. An estimation for IrMn and PdMn~\cite{ref:material1,ref:material2} is as follows: with $j_c\sim10^7$~A/cm$^2$, $\Delta\omega=\rho_s\mathcal{G}$ reaches $100$ GHz, which is comparable to the anisotropy gap $\omega_0$. Such an appreciable difference can be easily measured by AFM resonance~\cite{ref:AFMR}.

We also study the general case with finite $k$, and solve for $\omega(k)$ in the longitudinal direction. As current density is increased, the imaginary part of $\omega(k)$ changes sign at a threshold, beyond which damping turns into amplifying. As a result, spin waves at certain frequencies become unstable, \textit{i.e.}, magnons are emitted by fast moving electrons. The threshold spin current density is obtained by setting $\mathrm{Im}[\omega(k)]=0$ for a given $k$, 
\begin{align}
 j_s^{\mathrm{crit.}}=\frac{\omega_0c}{\mathcal{G}} \left[\frac{k}{k_0}+\frac{k_0}{k}\right], \label{eq:Cherenkov}
\end{align}
where $k_0=\omega_0/c$. Eq.~\eqref{eq:Cherenkov} reaches a minimum at $k=k_0$, which marks the most unstable mode. For this particular mode, the wave length is estimated to be $\lambda_0\sim 10^2$ nm for IrMn and PdMn~\cite{ref:material1,ref:material2}. Since $\lambda_0$ is much larger than the lattice spacing of the two materials, our assumption at the beginning is guaranteed.

For IrMn and PdMn, we also estimate that the threshold current density is of order $10^7$~A/cm$^2$. Again, this value will be much higher if the FM polarizer is not half metallic. But we stress that the instability solved above is a phenomenon peculiar to spin current injection. If the polarizer is completely removed, $\mathcal{G}$ will vanish and $j_s^{\mathrm{crit.}}$ will go to infinity, by which the instability will disappear. In fact, pure charge current leads to a Doppler shift of the spin wave velocity~\cite{ref:Duine,ref:MacD}; it is not able to trigger an instability of the same sense. Furthermore, it is remarkable that $\tilde{\alpha}$ does not appear in Eq.~\eqref{eq:Cherenkov}, though the instability is physically due to the overcoming of damping by the spin current.

We find that $\mathrm{Re}[\omega(k)]$ is also zero at the threshold point, which means the spin wave instability is not associated with propagating modes, but is in fact an instability towards the formation of \textit{stationary} a spatial pattern with period $2\pi/k_0$. When an inhomogeneous spatial configuration is developed, exchange energy of the AFM background is increased. Therefore, to sustain such a texture, energy of conduction electrons must be transferred continuously to the background moments. This may cause a sudden rise of the differential resistance $\mathrm{d}V/\mathrm{d}I$ at the threshold, which is detectable with high accuracy using today's technology~\cite{ref:SW2,ref:SW3}.

We thank E. Tveten, A. Brataas, A. Qaiumzadeh, J. Ieda, S. Maekawa, T. Ono, M. Tsoi, K. Everschor, A. Rosch, J. Xiao, O. Tretiakov, and A. MacDonald for helpful discussions. RC is grateful to G. E. W. Bauer for pointing out an important mistake in the first version, and to E. Saitoh for proposing experiments. This work is supported by DOE (DE-FG03-02ER45958), the MOST Project of China (2012CB921300), NSFC (91121004), and the Welch Foundation (F-1255).


\begin{thebibliography}{20}
  \bibitem{ref:spintronics} I. \v{Z}uti\'{c}, J. Fabian, and S. D. Sarma, Rev. Mod. Phys. \textbf{76}, 323 (2004) and the reference therein.
  \bibitem{ref:Volovik} G. E. Volovik, J. Phys. C \textbf{20}, L83 (1987).
  \bibitem{ref:BegerSlonczewski} L. Berger, Phys. Rev. B \textbf{54}, 9353 (1996); J. Slonczewki, J. Magn. Magn. Mater. \textbf{159}, L1 (1996).
  \bibitem{ref:Bazaliy} Y. B. Bazaliy, B. A. Jones, and S. -C. Zhang, Phys. Rev. B \textbf{57}, R3213 (1998).
  \bibitem{ref:STT} S. Zhang and Z. Li, Phys. Rev. Lett. \textbf{93}, 127204 (2004).
  \bibitem{ref:STTReview1} D. C. Ralph and M. D. Stiles, J. Magn. Magn. Mater. \textbf{320}, 1190 (2008).
  \bibitem{ref:Tserkovnyak} C. H. Wong and Y. Tserkovnyak, Phys. Rev. B \textbf{80}, 184411 (2009); Y. Tserkovnyak and C. H. Wong, Phys. Rev. B \textbf{79}, 014402 (2009).
  \bibitem{ref:STTReview2} A. Brataas, A. D. Kent, and H. Ohno, Nature Materials \textbf{11}, 372 (2012).
  \bibitem{ref:DWDynamics1} G. S. D. Beach, M. Tsoi, J. L. Erskine, J. Magn. Magn. Mater. \textbf{320}, 1272 (2008).
  \bibitem{ref:DWDynamics2} Y. Tserkovnyak, A. Brattas, and G. E. W. Bauer, J. Magn. Magn. Mater. \textbf{320}, 1282 (2008).
  \bibitem{ref:SW1} Z. Li and S. Zhang, Phys. Rev. Lett. \textbf{92}, 207203 (2004).
  \bibitem{ref:SW2} Y. Ji, C. L. Chien, and M. D. Stiles, Phys. Rev. Lett. \textbf{90}, 106601 (2003).
  \bibitem{ref:SW3} M. Tsoi \textit{et al.}, Phys. Rev. Lett. \textbf{80}, 4281 (1998); M. Tsoi, V. Tsoi, J. Bass, A. G. M. Jansen, and P. Wyder, Phys. Rev. Lett. \textbf{89}, 246803 (2002).
  \bibitem{ref:Tsoi} Z. Wei \emph{et al.}, Phys. Rev. Lett. \textbf{98}, 116603 (2007).
  \bibitem{ref:Spincurrent} S. Urazhdin and N. Anthony, Phys. Rev. Lett. \textbf{99}, 046602 (2007).
  \bibitem{ref:AFMSV} B. G. Park, \textit{et al.}, Nat. Mater. \textbf{10}, 347 (2011).
  \bibitem{ref:Numerical1} R. Wieser, E. Y. Vedmedenko, and R.~Wiesendanger, Phys. Rev. Lett. \textbf{106}, 067204 (2011).
  \bibitem{ref:Numerical2} Y. Xu, S. Wang, and K. Xia, Phys. Rev. Lett. \textbf{100}, 226602 (2008).
  \bibitem{ref:Numerical3} T. Jungwirth, \textit{et al.} Phys. Rev. B \textbf{83}, 035321 (2011).
   \bibitem{ref:AFMSpintronics} A. H. MacDonald and M. Tsoi, Phil. Trans. R. Soc. A \textbf{369}, 3098 (2011).
  \bibitem{ref:Brataas1} K. M. D. Hals, Y. Tserkovnyak, and A. Brataas, Phys. Rev. Lett. \textbf{106}, 107206 (2011).
  \bibitem{ref:Brataas2} E. G. Tveten, A. Qaiumzadeh, O. A. Tretiakov, and A. Brataas, Phys. Rev. Lett. \textbf{110}, 127208 (2013).
  \bibitem{ref:Duine} A. C. Swaving and R. A. Duine, J. Phys.: Cond. Mat. \textbf{24}, 024223 (2012); A. C. Swaving and R. A. Duine, Phys. Rev. B \textbf{83}, 054428 (2011).
  \bibitem{ref:MacD} P. M. Haney and A. H. MacDonald, Phys. Rev. Lett. \textbf{100}, 196801 (2008); A. S. N\'{u}\~{n}ez, R. A. Duine, P. Haney, and A. H. MacDonald, Phys. Rev. B \textbf{73}, 214426 (2006).
  \bibitem{ref:Kunitsyn} H. V. Gomonay, R. V. Kunitsyn, and V. M. Loktev, Phys. Rev. B, \textbf{85}, 134446 (2012); H. V. Gomonay and V. M. Loktev, Phys. Rev. B, \textbf{81}, 144427 (2010).
  \bibitem{ref:rancheng} R. Cheng and Q. Niu, Phys. Rev. B \textbf{86}, 245118 (2012).
  \bibitem{ref:note} Even the dynamics within a unit cell (internal dynamics) is itself adiabatic, which defines a geometric mapping from the $\bm{n}$-orbit to electron spin orbit (see Ref.~\cite{ref:rancheng}).
  \bibitem{ref:Haldane} F. D. M. Haldane, Phys. Rev. Lett. \textbf{50}, 1153 (1983); \emph{ibid}, \textbf{61}, 1029 (1988); E. Fradkin, \textit{Field Theories of Condensed Matter Systems} (Addison-Wesley, Reading, MA, 1991).
  \bibitem{ref:DiXiao} D. Xiao, M. Chang, and Q. Niu, Rev. Mod. Phys. \textbf{82}, 1959–2007 (2010).
  \bibitem{ref:SMF}  S. E. Barnes and S. Maekawa, Phys. Rev. Lett. \textbf{98}, 246601 (2007).
  \bibitem{ref:Shengyuan} S. A. Yang \emph{et al.}, Phys. Rev. Lett. \textbf{102}, 067201 (2009); S. A. Yang \emph{et al.}, Phys. Rev. B 82, 054410 (2010).
  \bibitem{ref:THE} M. Lee, W. Kang, Y. Onose, Y. Tokura, and N. P. Ong, Phys. Rev. Lett. \textbf{102}, 186601(2009); A. Neubauer \emph{et al.}, Phys. Rev. Lett. \textbf{102}, 186602 (2009); P. Bruno, V. K. Dugaev, and M. Taillefumier, Phys. Rev. Lett. \textbf{93}, 096806 (2004); J. Ye \emph{et al.}, Phys. Rev. Lett. \textbf{83}, 3737 (1999). 
  \bibitem{ref:supplementary} See Supplemental Material at \url{http://link.aps.org/supplemental/10.1103/PhysRevB.89.081105} for mathematical details and additional discussions.
  \bibitem{ref:spinpumping} Y. Tserkovnyak, A. Brataas, and G. E. W. Bauer, Phys. Rev. Lett. \textbf{88}, 117601 (2002).
  \bibitem{ref:photovoltaic} S. M. Young, F. Zheng, and A. M. Rappe, Phys. Rev. Lett. \textbf{110}, 057201 (2013).
  \bibitem{ref:material1} R. Y. Umetsu, M. Miyakawa, K. Fukamichi, and A. Sakuma, Phys. Rev. B \textbf{69}, 104411 (2004).
  \bibitem{ref:material2} S. Khmelevskyi, A. B. Shick, and P. Mohn, Phys. Rev. B \textbf{83}, 224419 (2011).
  \bibitem{ref:ExchBias} F. Nolting \emph{et al.}, Nature (London) \textbf{405}, 767 (2000); J. -V. Kim and R. L. Stamps, Phys. Rev. B \textbf{71}, 094405 (2005); D. Mauri, H. C. Siegmann, P. S. Bagus, and E. Kay, J. Appl. Phys. \textbf{62}, 3047 (1987).
  \bibitem{ref:DW} N. Papanicolaou, Phys. Rev. B \textbf{51}, 15062 (1995); \textit{ibid.} \textbf{55}, 12290 (1997).
  \bibitem{ref:Chien} F. Y. Yang and C. L. Chien, Phys. Rev. Lett. \textbf{85}, 2597 (2000).
  \bibitem{ref:Bode} M. Bode, E. Y. Vedmedenko, K. von Bergmann, A. Kubetzka, P. Ferriani, S. Heinze, and R. Wiesendanger, Nat. Mater. \textbf{5}, 477 (2006); M. Bode \emph{et al.}, Nature (London) \textbf{447}, 190 (2007); P. Sessi, N. P. Guisinger, J. R. Guest, and M. Bode, Phys. Rev. Lett. \textbf{103}, 167201 (2009).
  \bibitem{ref:Walker} N. L. Schryer and L. R. Walker, J. Appl. Phys. \textbf{45}, 5406 (1974).
  \bibitem{ref:AFMR} F. Keffer and C. Kittel, Phys. Rev. \textbf{85}, 329 (1952).
\end{thebibliography}
\end{document}